\documentclass[arguments]{aastex631}

\shorttitle{Leftover Planetesimals and MAB Implanted S-Complex SFDs}
\shortauthors{Deienno et al.}
\graphicspath{{./}{figures/}}

\begin{document}

\title{Size-Frequency Distribution of Terrestrial Leftover Planetesimals and S-complex Implanted Asteroids}

\correspondingauthor{Rogerio Deienno}
\email{rogerio.deienno@swri.org, rdeienno@boulder.swri.edu}

\author[0000-0001-6730-7857]{Rogerio Deienno}
\affiliation{Solar System Science \& Exploration Division, Southwest Research Institute, 1301 Walnut Street, Suite 400, Boulder, CO 80302, USA}

\author[0000-0003-1878-0634]{Andr\'e Izidoro}
\affiliation{Department of Earth, Environmental and Planetary Sciences, MS 126,  Rice University, Houston, TX 77005, USA}

\author[0000-0002-4547-4301]{David Nesvorn\'y}
\affiliation{Solar System Science \& Exploration Division, Southwest Research Institute, 1301 Walnut Street, Suite 400, Boulder, CO 80302, USA}

\author[0000-0002-1804-7814]{William F. Bottke}
\affiliation{Solar System Science \& Exploration Division, Southwest Research Institute, 1301 Walnut Street, Suite 400, Boulder, CO 80302, USA}

\author[0000-0001-7059-5116]{Fernando Roig}
\affiliation{Observat\'orio Nacional, Rua General Jos\'e Cristino 77, Rio de Janeiro, RJ 20921-400, Brazil}

\author[0000-0003-2548-3291]{Simone Marchi}
\affiliation{Solar System Science \& Exploration Division, Southwest Research Institute, 1301 Walnut Street, Suite 400, Boulder, CO 80302, USA}



\begin{abstract}

The isotopic composition of meteorites linked to S-complex asteroids has been used to suggest that these asteroids originated in the terrestrial planet's region, i.e., within 1.5 au, and later got implanted into the main asteroid belt (MAB). Dynamical models of planet formation support this view. Yet, it remains to be demonstrated whether the currently observed size-frequency distribution (SFD) of S-complex bodies in the MAB can be reproduced via this implantation process. Here we studied the evolution of the SFD of planetesimals during the accretion of terrestrial planets with the code LIPAD self-consistently accounting for growth and fragmentation of  planetesimals. In our simulations we vary the initial surface density of planetesimals, the gaseous disk lifetime, and the power slope of the initial planetesimals' SFD. We compared the final SFDs of leftover planetesimals in the terrestrial planet region  with the SFD of observed S-complex MAB objects (D $>$ 100km). We found that the SFDs of our planetesimal populations and that of S-complex MAB objects show very similar cumulative power index (i.e., q $\approx$ 3.15 in N($>$D)$~\propto$ D$^{-q}$) for slopes in the diameter range 100 km $<$ D $<$ 400 km by the end of our simulations. Our results support the hypothesis of S-complex MAB implantation from the terrestrial planet forming region, assuming implantation is size-independent, and implies that implantation efficiency is smaller than $\mathcal{O}$(10$^{\rm -2}$--10$^{\rm -4}$) to avoid over-implantation of (4) Vesta-sized objects or larger.

\end{abstract}

\keywords{Asteroid belt; Asteroids dynamics; Solar system formation; Terrestrial planets.}


\section{Introduction} \label{sec:intro}

The MAB is known for having a small total mass (${\rm \approx 5\times10^{-4}}$ Earth masses 0.05 lunar masses) and for being primarily composed of S- and C-complex taxonomic class bodies \citep[e.g.][]{Gradie1982,Mothe-Diniz2003,DeMeo2014}. Isotopic measurements from meteorites suggest C-complex bodies, many which are water/volatile-rich \citep[e.g.,][]{Robert1977,Burbine2002,Che2023}, are likely to have originated exterior to Jupiter’s orbit, whereas S-complex bodies, most which are relatively water/volatile-poor \citep[e.g.,][]{Kerridge1985,Burbine2002,Che2023}, come from the inner solar system \citep{DeMeo2014}.

The MAB’s low mass and taxonomic mixing\footnote{Characterized by the large dominance of S-complex bodies in the inner main belt, large C-complex bodies dominating the outer main belt, and a mix of both groups in the central main belt.} strongly constrain formation models. Some models suggest the MAB was originally much more massive than it is today and later depleted due to the effects of a potential early giant planet migration \citep{Walsh2011} or orbital instabilities \citep{Clement2019}. A key caveat against models invoking a massive MAB  was pointed out in a new study  by \citet{Deienno2024} who demonstrated that the primordial MAB mass had to be no larger than $\approx$ 4 times its current mass. A relatively massive MAB would have led to the formation of many large (D $>$ 500 km) bodies in MAB that are not observed today and would have likely survived any subsequent depletion over the history of the solar system \citep{Deienno2024}. In contrast, the idea that the MAB was originally empty \citep{Raymond2017a,Raymond2017b} or had a low initial mass \citep{Izidoro2015,Johansen2015,Deienno2023,Deienno2024,Brasser2025}  remains a viable and appealing hypothesis given its present-day low mass.

In this view, the modern MAB would have formed primarily through the limited implantation of S- and C-complex asteroids that were scattered from their different source regions, i.e., the inner and outer solar system \citep{Raymond2017a,Raymond2017b,Deienno2022,Deienno2023,Deienno2024,Marchi2022,Nesvorny2024,Avdellidou2024,Izidoro2024,Izidoro2024Athor}. C-complex bodies are thought to have been scattered inwards and implanted predominantly in the central and outer MAB early in the solar system's history, during the giant planet growth phase, when gas was still present in the protoplanetary disk. As Jupiter/Saturn grew and migrated, it would have scattered some planetesimals inward toward the Sun. Gas drag then acted to remove these objects from giant planets' crossing orbits, allowing some of them to become trapped in the MAB.

The terrestrial planets (having relatively low masses compared to growing gas giant planets) were not capable of gravitationally  scattering objects from the inner solar system regions ($a<$ 1.5~au) into the MAB during the nebular gas phase. This holds true even if scattering occurred near the very end of the gas nebula phase, when the gas density was relatively low and its damping effect on scattered objects was potentially weaker. Therefore, unless considering the possibility of large-scale radial gas-driven migration of Jupiter and Saturn through the inner solar system \citep{Walsh2011}, the implantation of S-complex objects from the terrestrial planet region likely occurred after the dissipation of the solar nebula, and it may have been facilitated by additional processes as the giant planets' orbital instability \citep{Deienno2018,Clement2019,Avdellidou2024,Izidoro2024}. The bottom line is that, if we consider that implantation of S-complex bodies into the MAB was indeed necessary  \citep[e.g., in case the MAB was originally low mass, or even empty;][]{Deienno2024} and that Jupiter and Saturn never coursed through the inner solar system \citep{Walsh2011}, such implantation occurred relatively later in time than those of C-complex bodies. 
 
The initially empty or low mass MAB hypothesis is supported by recent models suggesting our solar system formed out off concentric rings of planetesimals at various radial distances around the Sun \citep{Lichtenberg2021,Izidoro2022,Morbidelli2022,Nesvorny2024}.  These models are motivated in part by the distinct  isotopic compositions of non-carbonaceous (NC) and carbonaceous (CC) solar system materials  \citep[e.g.,][the so-called NC-CC isotopic dichotomy]{Warren2011,Kruijer2017,Nanne2019,Burkhardt2021}. Under this framework, some  models suggest that the terrestrial planets formed from a ring of material centered at 1 AU~\citep{Hansen2009,Izidoro2022,Morbidelli2022} and that the S-complex population in the MAB samples material from this region~\citep{Raymond2017b,Izidoro2024}.

\begin{figure}[!ht]
    \centering
    \includegraphics[width=13.cm]{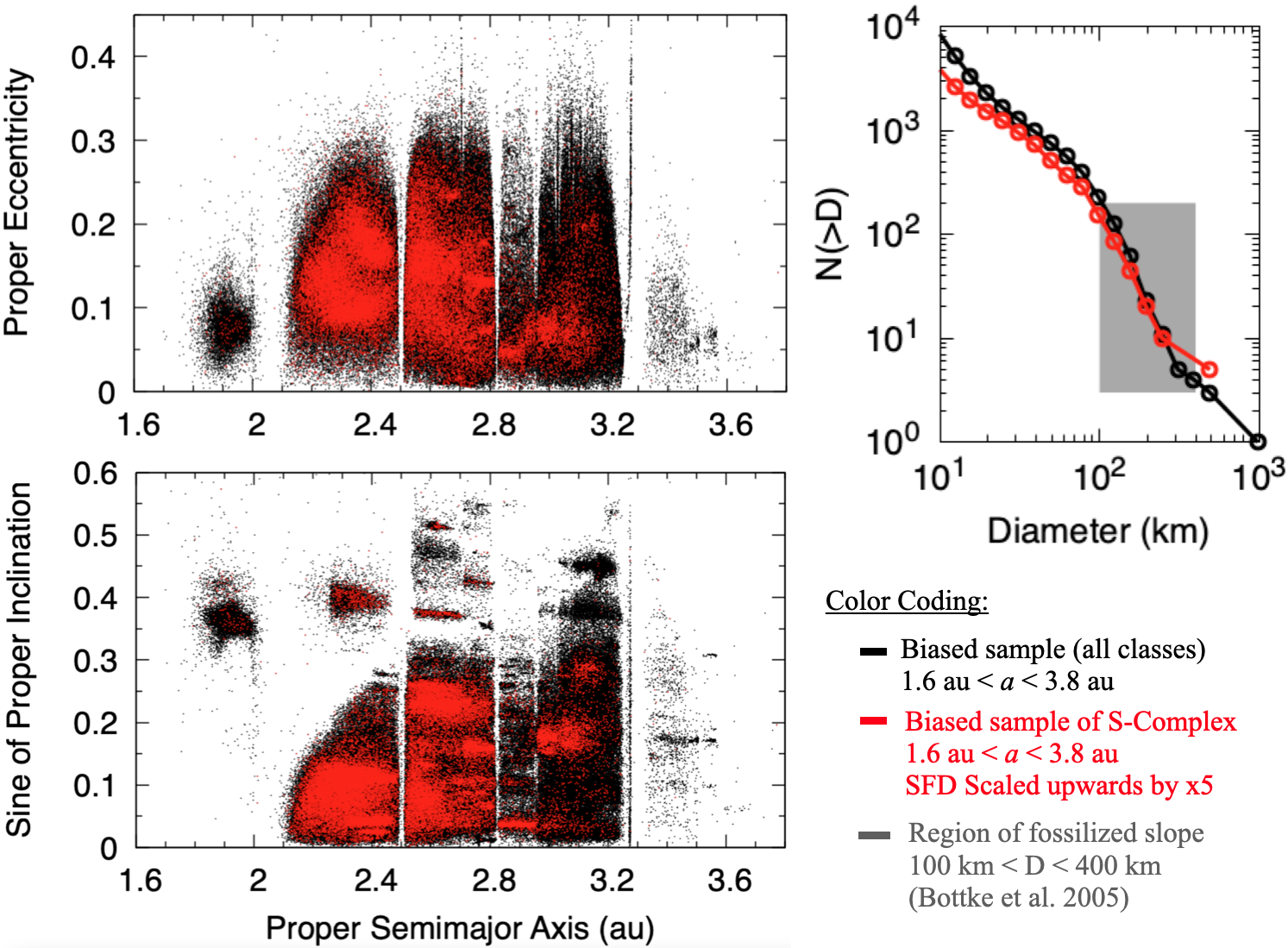}
    \caption{Comparison between known asteroids of S-complex taxonomical classification, i.e. {\it S, Sq, Sk, Q, Sr, V}, which includes asteroid (4) Vesta (red dots), and those from all classes {\it S, Sq, Sk, Q, Sr, V, B, C, Cb, Ch, Cg, Cgh, X, Xk, Xc, Xe, E, M, P} including S-complexes \citep[black dots;][see also \citet{Delbo2017,Delbo2017db},  and more recently \citet{DeMeo2022}]{Delbo2019}. Top left panel compares proper eccentricity with proper semimajor axis. Bottom left panel compares sine of proper inclination with proper semimajor axis. Top right panel compares the cumulative SFD of the two populations. The S-complex SFD (red symbols in the top right panel) was scaled upwards by a factor of 5. Gray shaded area delimits the diameter range (100 km $<$ D $<$ 400 km) where the SFD power slope was probably unchanged over solar system history \citep{Bottke2005}. For the reference, a cumulative power slope q $\approx$ 3.15 in N($>$D) $ \propto$ D$^{-q}$ is representative of the S-complex SFD in this size range. A color coding caption is also added to the bottom right of the figure to aid interpretation.}
    \label{Fig1}
\end{figure}

An open question arising from these scenarios is whether the SFD of the MAB components can be explained either by the implantation \citep[e.g.,][]{Walsh2011,Raymond2017b,Izidoro2024} or local collisional growth \citep[e.g.,][]{Carter2015,Clement2020gpu} of S- and C-complex asteroids. Here we focus on the former. Existing simulations modeling the implantation of planetesimals into the MAB have generally overlooked the effects of imperfect accretion and the long-term collisional evolution of planetesimals when comparing the implanted population to present-day observations. Collisional modeling work shows that the power law slope of the MAB’s SFD for objects 100 km $\leq$ D $\leq$ 400 km in diameter does not meaningfully change over solar system history \citep[see also \citet{Bottke2005b,Morbidelli2009,Bottke2020}]{Bottke2005}. Based on that, \citet{Bottke2005} concluded that the power law slope in this size range is a fossil of that population's primordial SFD. This result implies that implanted asteroids in this size range should have a slope comparable to the MAB’s current slope (Figure \ref{Fig1}).

\citet{Deienno2022} used this constraint to propose the following. Consider that C-complex asteroids \citep {DeMeo2014,Bottke2020} formed as planetesimals between Jupiter's original orbit and 5 au \citep[e.g.][]{Bitsch2015a,Johansen2017}. That would imply that the SFD of planetesimals that formed in the Jupiter zone either had a primordial shape that resembles the one observed in the current MAB in the 100 km $\leq$ D $\leq$ 400 km diameter range or collisionally evolved to it prior to implantation. Furthermore, to avoid excessive mass implantation into the MAB, \citet{Deienno2022} suggested that Jupiter likely did not form beyond 10--15 au from the Sun.

More recently, \citet{RibeiroDeSousa2024} reassessed the problem of gas-assisted C-complex asteroid implantation.  Their model accounted for how surface mass ablation \citep{Eriksson2021} may have affected the SFD of the implanted population \citep[see also][]{Nesvorny2024}. The authors found that small and large water-rich C-complex objects ablate at the same rate if they have the same eccentric orbit, but larger objects take longer to circularize by gas drag. As a consequence, mass ablation tends to produce an implanted C-complex SFD that is slightly steeper than that of the initial assumed SFD, where the absolute value of the difference between the power slopes for the implanted and original SFDs is smaller than unity.

In this work we focus on the implantation of S-complex asteroids into the MAB from the terrestrial planet region \citep{Raymond2017b,Avdellidou2024,Izidoro2024}, while assuming the MAB was primordially never massive, as needed \citep[see also \citet{Brasser2025}]{Deienno2024} to avoid the growth of unobserved large, D $>$ 500 km, objects in that region that would survive posterior depletion \citep{Clement2019}. 

Figure \ref{Fig1} compares the proper semimajor axis, proper eccentricity, and sine of proper inclination for all asteroids in the semimajor axis range between 1.6 au $<a<$ 3.8 au with known taxonomical designation (all classes; black) with those of S-complex\footnote{We retrieved proper elements, diameters, and taxonomical classification for our asteroid sample from \url{https://mp3c.oca.eu} via `Old Best Values' under the `Search' drop-down menu and following instructions included in `Search by parameters' tab. We restricted our search to proper semimajor axes between 1.6 au $<a<$ 3.8 au while considering the entire proper eccentricity and sine of proper inclination ranges, and selecting `Spectral classes' as follows: (1) {\it S, Sq, Sk, Q, Sr, V, B, C, Cb, Ch, Cg, Cgh, X, Xk, Xc, Xe, E, M, P} for all classes, i.e., black in Figure \ref{Fig1}, and (2) {\it S, Sq, Sk, Q, Sr, V}, which includes asteroid (4) Vesta, for S-complex only, i.e., red in Figure \ref{Fig1} \citep[][see also \citet{Delbo2017,Delbo2017db},  and more recently \citet{DeMeo2022}]{Delbo2019}.} alone \citep[red;][see also \citet{Delbo2017,Delbo2017db},  and more recently \citet{DeMeo2022}]{Delbo2019}. 
Figure \ref{Fig1} also compares the individual SFDs of those populations  (right panel). As previously reported \citep[see also \citet{DeMeo2022}]{DeMeo2014}, S-complex asteroids have a strong presence in the inner MAB regions. More importantly for the scope of our work is that the power slope of the two SFDs, S-complex (red) and all classes (black), are similar between 100 km $<$ D $<$ 400 km (gray shaded region). The fact that they have the same shape is intriguing. Recall that the MAB population is dominated in both number and mass by C-complex objects \citep{DeMeo2014}, and these objects come from the giant planet zone, while S-complex bodies (SFD scaled upwards by a factor of 5 in Figure \ref{Fig1}) come from the inner solar system. This could suggest similar planetesimal formation processes and/or similar degrees of collisional evolution prior to implantation in the MAB. Once in the MAB, collisional models show that the power slope in this size range probably did not change over solar system history \citep{Bottke2005}. 

A related constraint for the proposed implantation scenario for S-complex bodies \citep{Raymond2017b,Izidoro2024} is that they must reproduce this power law slope in their evolution models of leftover planetesimals from the terrestrial planet formation. After solar nebula dispersal and during late stage accretion, the evolved terrestrial planet leftover planetesimal SFD should mimic the shape of the SFD in red within the gray shaded area in Figure \ref{Fig1} (i.e., the current MAB S-complex SFD). 

The above constraint is valid unless implantation from the terrestrial planet region into the MAB is size-dependent. It is not clear, however, why size sorting would be expected especially if the objects need to be captured in the MAB after the gas disk is gone \citep{Avdellidou2024,Izidoro2024,Izidoro2024Athor}. Testing the potential effects of size sorting during implantation is beyond the scope of this paper. While modeling post-nebula implantation mechanisms from the terrestrial planet region would be illuminating, they would also be computationally expensive, with the runs requiring (i) extremely large numbers of planetesimals to glean insights into capture processes and (ii) multiple case studies, with the results being sensitive to planetary embryo evolution after the end of the nebula \citep{Raymond2017b,Izidoro2024,Izidoro2024Athor,Avdellidou2024}. We leave this to future work. 

Our investigation in the present work will be based solely on comparing the SFD of leftover planetesimals formed in the terrestrial planet region, following the framework by \citep{Izidoro2022}, at the time of gas disk dispersal with the current S-complex SFD observed in the MAB for diameters between 100 km $<$ D $<$ 400 km (red within gray area in Figure \ref{Fig1} right panel). We will not consider mass ablation effects of S-complex in our modeling of the SFD of leftover planetesimals during the solar nebula phase. \citet{RibeiroDeSousa2024} showed that mass ablation is very limited or even inexistent for Enstatite-type (S-complex) bodies, unless very close to the Sun ($a\ll$ 0.7--1 au) where disk temperatures are elevated.

\section{Methods} \label{sec:methods}

\begin{figure}[!ht]
    \centering
    \includegraphics[width=\linewidth]{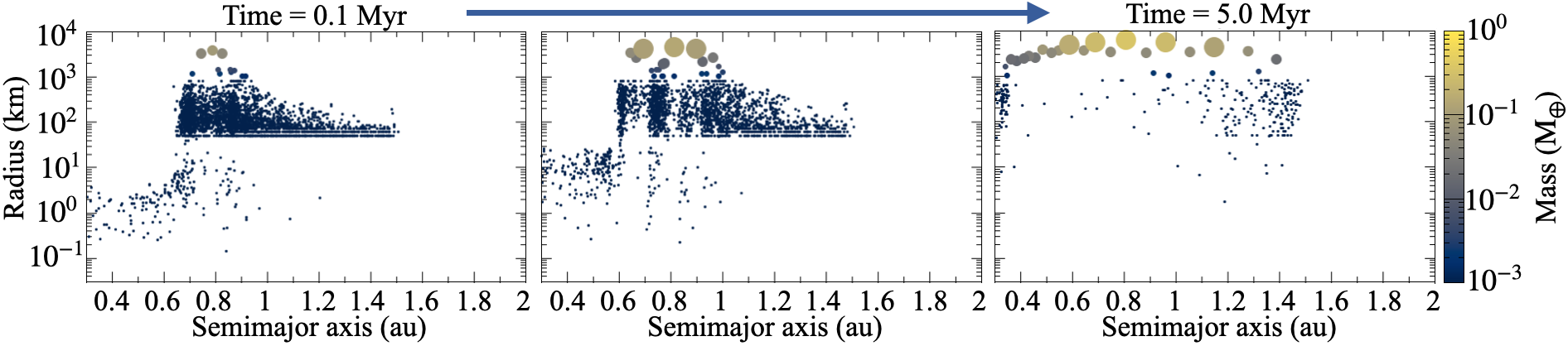}
    \caption{Representative evolution of terrestrial planet accretion starting from a planetesimal distribution as conducted in this work using LIPAD. This specific evolution (see main text for details in simulation parameters) assumes ${\rm M_{disk} \approx}$ 2.5 Earth masses distributed into 3000 tracer particles within 0.7 au $\leq a \leq$ 1.5 au. Planetesimals started from D$_{\rm 0}$ = 100 km (i.e., initial SFD power slope $q$ = 0) and followed a surface density proportional to $\gamma=$ 5.5. The nebula gas disk exponentially decays with ${\rm \tau_{gas}}=$ 2 Myr. For additional reference, these are the exact same parameters adopted in one of the \citet{Izidoro2022} representative cases (see their Supplementary Information Figure 7).}
    \label{Fig2}
\end{figure}

We conducted terrestrial planet formation simulations starting from planetesimal-sized objects that were tracked over 5 Myr within the solar nebula (Figure \ref{Fig2}). Our code assumed the gas disk exponentially dispersed over timescales $\tau_{\rm gas}$ of 0.5, 1, and 2 Myr \citep{Haisch2001}. The exact time of gas nebula dispersal in the protoplanetary disk is unknown, but chondrules (i.e., igneous spherules found in meteorites) have relative formation ages to Calcium-Aluminium-Inclusions (CAIs) of about 3 Myr \citep[e.g., ordinary chondrites;][]{Pape2019} to 5 Myr \citep[e.g., CB chondrites;][]{Krot2005}. It is thought that chondrules can only form and get into planetesimals while gas in the protoplanetary disk exists \citep{Johnson2016}. Therefore, the age of the youngest (last) formed chondrules are indicative of the protoplanetary gas disk lifetime. Furthermore, those ages are also in good agreement with paleo-magnetism constrains, suggesting that the nebular gas dispersal time in the inner and outer solar system was approximately 1.22 Myr $<$ T$_{\rm inner}^{\rm gas}$ $<$ 3.94 Myr and 2.51 Myr $<$ T$_{\rm outer}^{\rm gas}$ $<$ 4.89 Myr \citep{Weiss2021}. 

We gave our initial planetesimal population a cumulative SFD following ${\rm N(>D) \propto D^{-q}}$. Several streaming instability calculations can successfully reproduce what is expected to be the primordial SFD of large Kuiper belt objects \cite[e.g.,][]{Youdin2005}, finding a power slope $q$ around $\approx$ 5 for objects with D $>$ 100 km a good fit. This value is in agreement with current observations of that population \citep[e.g.][]{Fraser2014} and with expectations from dynamical \citep[e.g.][]{Nesvorny2018a} and collisional modeling \citep[e.g][]{Bottke2023,Bottke2024sat}.

Unfortunately, for planetesimals in the terrestrial planet forming region, little is known about that population's primordial SFD \citep[e.g.][]{Johansen2015,Simon2016}.
For this reason, it is common for terrestial planet formation models to start from either a population of large (Moon- to Mars-sized) embryos \citep[e.g.][]{Hansen2009,Izidoro2015,Clement2018,Nesvorny2021} or from a population of equal, or nearly equal, size planetesimals \citep[e.g.][]{Walsh2016,Walsh2019,Deienno2019,Clement2020,Woo2023}.  For the latter, we represent them using $q = 0$ and a reference diameter D$_{\rm 0}$ in ${\rm N(>D) = D_{0}D^{-q}}$.

Alternative models assume that objects in the terrestrial region should have a SFD power slope similar to that of the current MAB in the diameter range of 100 km $<$ D $<$ 1000 km \citep[e.g.][]{Levison2015b}. If so, this power law slope can be loosely approximated by $q \approx 3.5$. 

In this work, we assume all three values are equally plausible. Thus, we attribute $q = 0$ with $D_{\rm 0} = 100$~km \citep{Klahr2020,Klahr2021}, as well as $q = 3.5$ \citep{Levison2015b} and $q = 5$ \citep[e.g.][]{Youdin2005} for diameters in the range $100 < D < 1000$~km for our initial planetesimal SFD. 

Following planetesimal formation simulations by \cite{Izidoro2022}, we adopted two initial planetesimal surface density distributions. Planetesimals are initially radially distributed according to ${\rm \Sigma_{planetesimal} \propto r^{-\gamma}}$ with $\gamma =$ 1 or 5.5 in the semi-major axis range $0.7 \leq a \leq$~1.5 au. The exact value of $\gamma$ is dependent, for instance, on the disk evolution and planetesimal formation model considered \citep[e.g.,][]{Drazkowska2018,Morbidelli2022}. The values of $\gamma =$ 1 and 5.5 represent two end-member cases from simulations by \citet{Izidoro2022}. The total mass considered in all of our simulations is $\approx 2.5$~Earth masses \citep{Izidoro2022}. 

We follow the growth of the terrestrial planets with the code LIPAD \citep{Levison2012} (Figure \ref{Fig2}). LIPAD is a well-tested code capable of self-consistently handling accretion, growth, and fragmentation of sub-kilometer objects (represented by tracer particles) that are in the process of becoming planets \citep[e.g.,][]{Walsh2016,Walsh2019,Deienno2019}. Through its lagrangian and statistical approaches \citep{Levison2012}, LIPAD is to our knowledge the only N-body code capable of resolving collisional fragmentation and dynamics of several million self-gravitating particles, thus allowing for tracking SFD evolution with enough resolution, and making it ideally suited to the task. The evolution of the planetesimal SFD is monitored throughout the accretion simulation. Some bodies eventually grow to become planetary embryos (see large circles above $\approx$ 1000 km in radius in Figure \ref{Fig2}). We have used 3000 tracer particles as the initial tracer population in our simulations \citep{Walsh2019,Izidoro2022}.

Assuming that S-complex planetesimals are implanted in the MAB after the solar nebula has dispersed \citep{Raymond2017b,Izidoro2024}, once the simulations are complete, comparisons are made between our evolved SFD at the end of the simulation (e.g., small dots below $\approx$ 1000 km in radius in Figure \ref{Fig2}) with that from S-complex asteroids in the current MAB (red in Figure \ref{Fig1}) in the range $100 < D < 400$~km. We also extended two representative simulations from 5 Myr (end of gas disk phase) to 100 Myr in order to account for potential changes in the power law slope of the planetesimal SFD during late stage accretion and the MAB implantation epoch.

\section{Results} \label{sec:results}

\begin{figure}[!ht]
    \centering
    \includegraphics[width=0.85\linewidth]{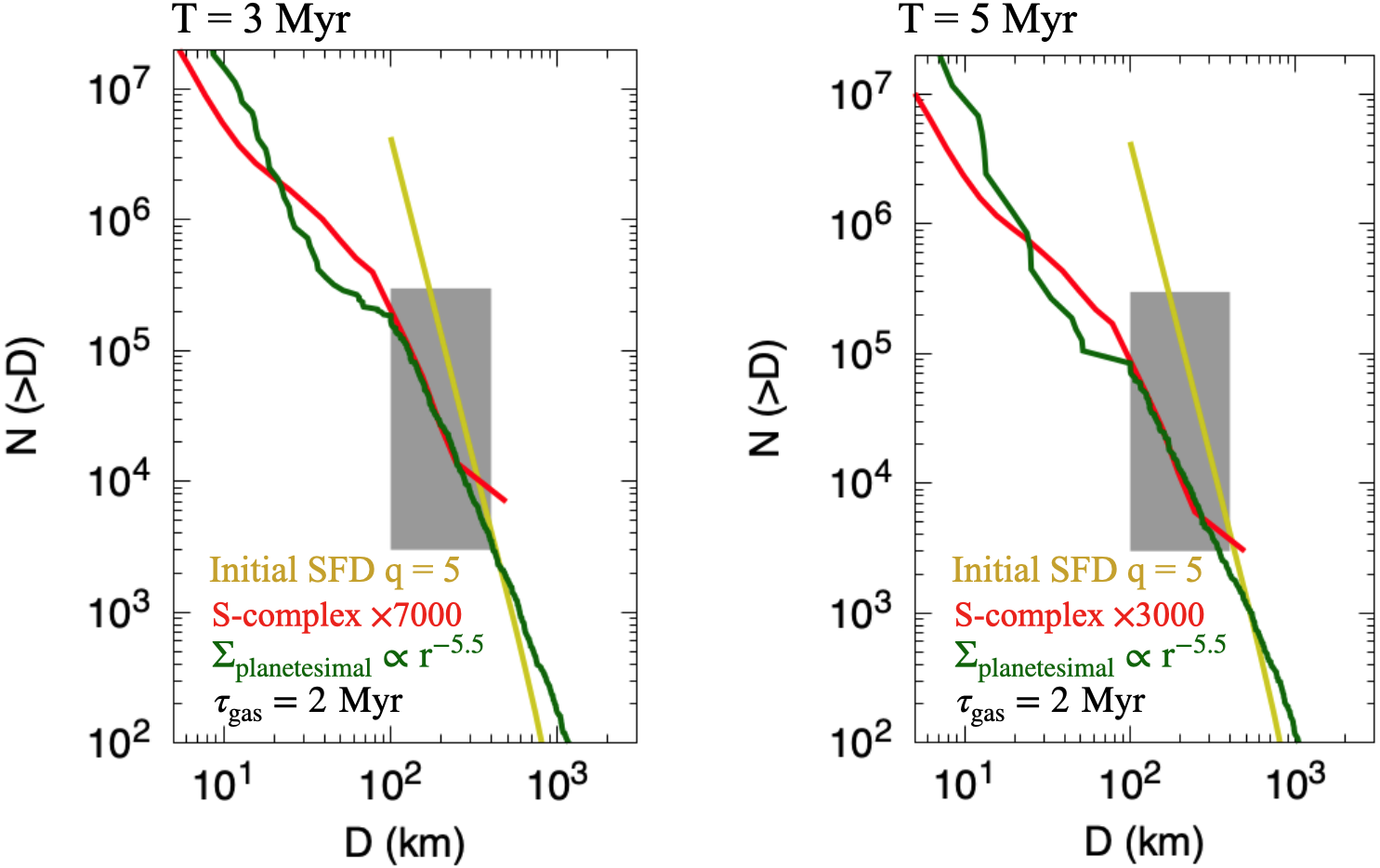}
    \caption{Comparison between leftover planetesimals evolved SFD in the terrestrial planet region (green) — for ${\rm \tau_{gas}}=$ 2 Myr,  $q = 5$, and $\gamma=$ 5.5  — with the scaled MAB S-complex SFD (red). The yellow line is shown for reference of the initial SFD. The gray shaded area delimits the diameter range where the leftover planetesimal evolved SFD slopes should be compared to that of the S-complex SFD (Section \ref{sec:intro}, Figure \ref{Fig1}). Left panel shows the result for T = 3 Myr after CAIs. Right panel is for T = 5 Myr after CAIs.}
    \label{Fig3}
\end{figure}

We have followed a total of 18 simulations ($\gamma=$ 1 and 5.5, ${\rm \tau_{gas}}=$ 0.5, 1, and 2 Myr, with $q = 0$, 3.5 and 5; Section \ref{sec:methods}) of the accretion of planetesimals to proto-planets for 5 Myr, from an initial disk mass equal to $\approx$ 2.5 Earth masses (see Figure \ref{Fig2} for snapshots of a representative evolution\footnote{Although not necessarily important for the goals of the present work (which focuses on the period when gas disk is still present), it is worthy noticing that the distribution of planetesimals and protoplanets presented in Figure \ref{Fig2} are similar, if not identical, to previous works conditions at the end of the gas disk phase that reproduced the terrestrial planet system with some success after 100-200 Myr \citep[e.g.,][to cite a few.]{chambers01,Raymond2009,Jacobson2014,Walsh2016,Walsh2019,Izidoro2015,Izidoro2022,Clement2018,Clement2020gpu,Deienno2019,Nesvorny2021}.}).
Given our uncertainty regarding the time when the solar nebula gas fully dispersed in the terrestrial planet region \citep{Weiss2021}, as discussed in Section \ref{sec:methods}, we report our results on the evolved SFDs for both T = 3 and 5 Myr after CAIs.
In Figure \ref{Fig3} we show the comparison of our leftover planetesimal evolved SFD in the terrestrial planet region with the current MAB S-complex SFD for one representative case assuming ${\rm \tau_{gas}}=$ 2 Myr,  $q$ = 5, and $\gamma=$ 5.5. 
A compilation of the results showing all other evolved leftover planetesimal SFDs from all 18 simulations considering different values of ${\rm \tau_{gas}}$, $q$, and $\gamma$, is presented in Figure \ref{Fig3Apx} in the Appendix Section \ref{sec:apx}. Our SFDs only account for objects smaller than $\approx$ 1000 km, i.e., growing proto-planets (D $\gtrsim$ 1000 km) are not included. 

A large concentration of mass is needed near 1 au to form the terrestrial planets and reproduce their orbits \citep{Hansen2009,Jacobson2014,Izidoro2022}. This causes collisional evolution there to be intense. As a result, the planetesimal SFD rapidly reaches collisional equilibrium \citep[T $\ll$ 3 Myr;][]{Bottke2007}. This is true in all simulations, for all evolved planetesimals SFDs, regardless of $q$, $\tau_{\rm gas}$, and $\gamma$. 

Once in equilibrium, the slope of all of our evolved leftover planetesimal SFDs (see Figure \ref{Fig3Apx}) reasonably match that of the current MAB S-complex SFD in the range 100 km $\leq$ D $\leq$ 400 km (slopes within the gray shaded area in Figures  \ref{Fig3} and \ref{Fig3Apx}). This similar shape broadly satisfies the constraint defined in Sections \ref{sec:intro} (Figure \ref{Fig1} right panel) and \ref{sec:methods}. Indeed, Figure \ref{Fig3} shows a good agreement between the evolved leftover planetesimal SFD (green) and that of the S-complex MAB (red). 

Figure \ref{Fig3Apx}, on the other hand, shows some small differences (i.e., some of the evolved slopes of the SFD may end up slightly shallower than the observed slope of the S-complex SFD, while others may end up slightly steeper).  It is useful to consider, however, that our S-complex SFD slope was obtained from a biased sample of limited asteroids with known taxonomic classification (see Section \ref{sec:intro} and Figure \ref{Fig1}). Therefore, given the similarities of all SFD slopes within the gray region in Figure \ref{Fig3Apx} (and specially in Figure \ref{Fig3}), it is reasonable to consider that such differences are minor, and that the overall agreement between our evolved SFDs with the one observed for S-complex asteroids is good.

The fact that, regardless the initial values of $q$, $\tau_{\rm gas}$, and $\gamma$ used, they all evolve to the same SFD slope around 100 km $<$ D $<$ 400 km range, as shown in Figure \ref{Fig3Apx} servers as an indicative that testing other initial values in between for those quantities would essentially lead to the same results. This happens because while gas is still present in the nebula (T $\lesssim$ 5 Myr) planetesimals tend to have their orbits damped to low eccentricities and inclinations. In this regime, the way that planetesimals collisionally evolve and grow into protoplanets does not produce results that are meaningfully different between different runs. Therefore, our 18 simulations are sufficient to our goal in this work, i.e, they produced qualitatively similar results, demonstrating their robustness for capturing the overall expected evolution of the SFD of terrestrial planet leftover planetesimals.

The evolved SFDs shown in Figures \ref{Fig3} and \ref{Fig3Apx} represent the leftover planetesimals that will be available for implantation into the MAB. Those SFDs are also what we should expect to be representative of the leftover planetesimal population available for generating craters on terrestrial worlds and the Moon from early bombardment \citep[e.g.,][among many others]{Hartmann1966,Hartmann2001,Neukum2001,Marchi2009,Neumann2015,Minton2015,Morbidelli2018,Brasser2020crater,Robbins2022}.  

More recently, while addressing the findings from all those previous works alongside with results from a very successful terrestrial planet formation simulation \citep{Nesvorny2021}, \citet{Nesvorny2022,Nesvorny2023} was able to satisfy several constrains based on both the lunar and martian crater chronology. Here, we use their work as the basis for comparison with ours and to discuss implications. 

\citet{Nesvorny2022,Nesvorny2023} considered an initial leftover planetesimal SFD power slope $q = 5$ for $D > 100$~km at the end of the gas disk phase, once protoplanets had already formed, while in this work we considered three different values of $q=$ 0 (D$_0$ = 100 km), 3.5 and 5 as the initial conditions at the onset of planetesimal formation. As shown in Figures \ref{Fig3} and \ref{Fig3Apx}, at the time of gas disk dispersal (the beginning of the \citet{Nesvorny2022,Nesvorny2023} simulations), regardless of what primordial SFD power slope we considered, they all should have substantially evolved towards an equilibrium. Although in reality the initial conditions used by \citet{Nesvorny2022,Nesvorny2023} should be different from what they considered, \citet{Nesvorny2022,Nesvorny2023} reports that after $\approx$ 20 Myr of collisional evolution, their initial SFD power slope $q = 5$ evolves to a MAB-like shaped SFD for $D > 100$~km. We argue that our evolved leftover planetesimal SFDs look similar to that shape in Figures \ref{Fig3} and \ref{Fig3Apx}. We attribute the differences between \citet{Nesvorny2022,Nesvorny2023} and our work to the fact that we self-consistently modeled the evolution of the planetesimal SFD during terrestrial planet accretion, whereas the choices made in \citet{Nesvorny2022,Nesvorny2023} were ad hoc at the time of the solar nebula dispersal. 

Our results show that the equilibrium state \citet{Nesvorny2022,Nesvorny2023} found after $\approx$ 20 Myr (after gas disk dispersal) should actually be reached at earlier times (T $\ll$ 3 Myr, while nebular gas is still around). 
These differences, however, are small enough to meaningfully affect results, particularly because early lunar cratering had to deal with (i) the potential solidification of the Lunar Magma Ocean \citep[T $\gtrsim$ 200 Myr; e.g.][]{Meyer2010,Elkins-Tanton2011,Morbidelli2018}, or (ii) a potential lunar remelting at around 4.35 billion years ago that was driven by tidal evolution \citep{Nimmo2024}. This means that, any large lunar basin that formed before the complete crystallization of the Moon's crust may be unidentifiable today \citep{Miljkovic2021}. 

The question that remains is whether the leftover planetesimal SFDs change their shape during the late stages of accretion and the early bombardment era. \citet{Nesvorny2022,Nesvorny2023} noticed that in their simulations, after the leftover planetesimals evolved into a MAB-like SFD, the population had become so decimated by collisional evolution that the subsequent collisional evolution was inconsequential. Their results verified what \citet{Bottke2007} have found in their investigation of what happens to leftover planetesimals.   

\begin{figure}[!ht]
    \centering
    \includegraphics[width=0.85\linewidth]{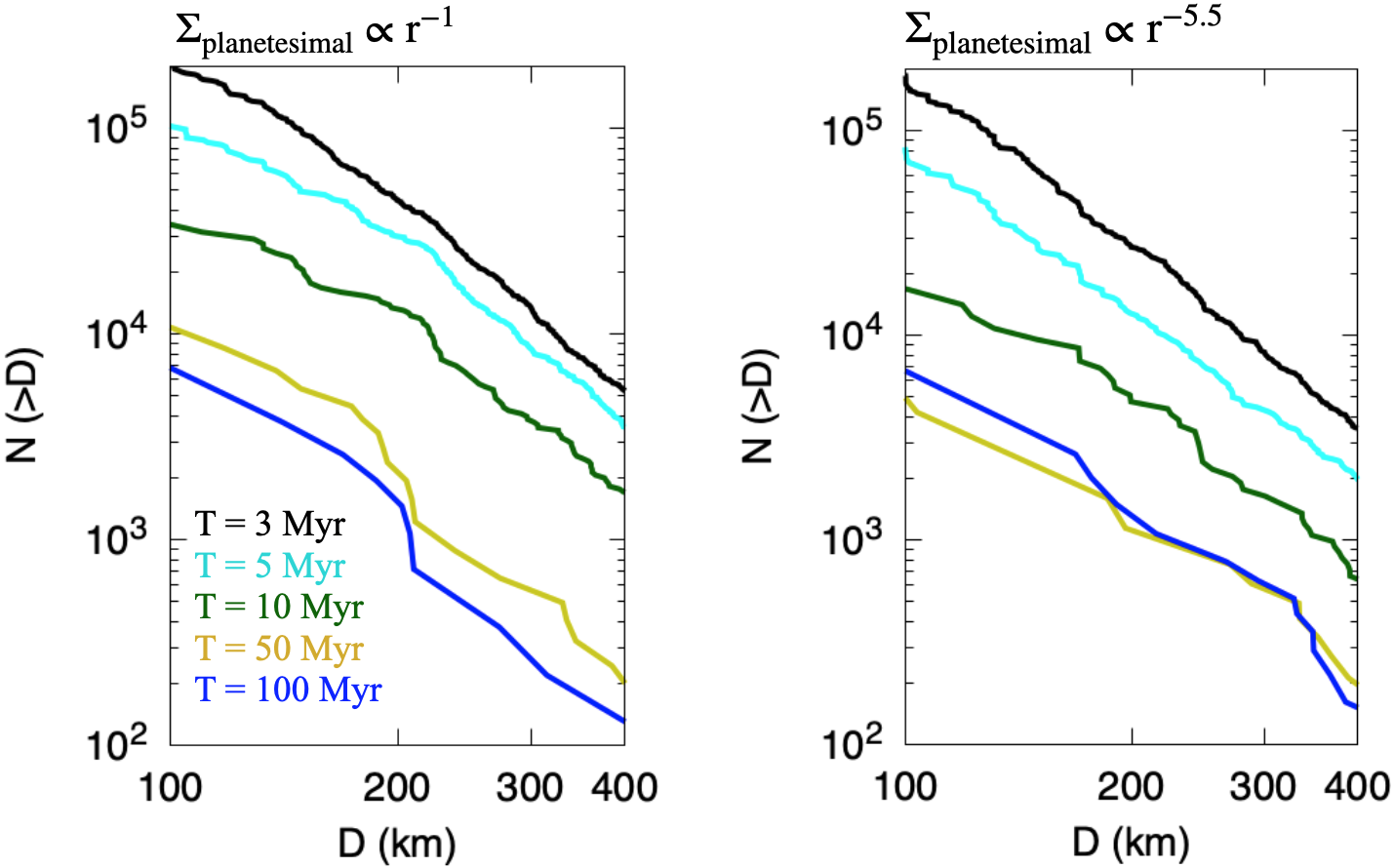}
    \caption{Evolution of  leftover planetesimal SFDs in the terrestrial planet region after gas disk dispersal. Colors show the different SFDs at T = 3 Myr (black) and 5 Myr (turquoise), with both times still in the gas disk phase (Figures \ref{Fig3} and \ref{Fig3Apx}), T = 10 Myr (green), 50 Myr (yellow), and 100 Myr (blue). We show the diameter range $100 < D < 400$~km where the slopes of the leftover planetesimal SFDs can be compared to that of the S-complex SFD (red in Figures \ref{Fig3} and \ref{Fig3Apx}, or black and turquoise in this Figure \ref{Fig4}).  Ideally, the model runs should keep a nearly constant slope over time. The black and turquoise lines for $T = 3$ and 5 Myr after CAIs are repeated here for comparison with the SFD at the time of gas disk dispersal from Figures \ref{Fig3} and \ref{Fig3Apx}. The left panel is for $\gamma=$ 1, while the right panel is for $\gamma= 5.5$. Both simulations are from cases assuming ${\rm \tau_{gas}}= 2$~Myr and $q = 5$ as initial conditions.}
    \label{Fig4}
\end{figure}

To confirm the reported findings, and better understand how the SFD of leftover planetesimals evolves during late stage terrestrial planet accretion \citep[and S-complex MAB implantation;][]{Raymond2017b,Izidoro2024,Izidoro2024Athor,Avdellidou2024}, we extended two of our simulations shown in Figure \ref{Fig3Apx} to T = 100 Myr. Those two cases are shown in Figure \ref{Fig4}, and they refer to the ones assuming as inital conditions: ${\rm \tau_{gas}=}$ 2 Myr and $q$ = 5, with $\gamma=$ 1  for the left panel, and 5.5 for the right panel. Recall that our simulations self-consistently account for both accretion and fragmentation, and that according to our results the specific choice of initial conditions should not affect conclusions as they all evolve to the same SFD shape for 100 km $<$ D $<$ 400 km at the time of gas nebula dispersal (Figure \ref{Fig3Apx}).

Results from Figure \ref{Fig4} confirm the findings of \citet{Bottke2007} and \citet{Nesvorny2022,Nesvorny2023}, namely that the overall slope of the leftover planetesimal SFD after gas disk dispersal is nearly unchanged in the $100 < D < 400$~km size range.  This implies that limited collisional evolution took place because the remnant population was relatively small after having been ground down by comminution. The latter action is a well-known result from terrestrial planet formation models \citep[e.g.][]{chambers01,Jacobson2014,Clement2018,Clement2020gpu,Deienno2019,Nesvorny2021,Woo2023,Brasser2025}.  

One issue in our simulations is that the fast decrease in the number of tracer particles leads to graininess in the resolution of our evolved SFDs for large $T$ values (see the large wiggles in the yellow and blue curves in the left panel of Figure \ref{Fig4} for T = 50 and 100 Myr). Another potential issue is that once the gas nebula has fully dispersed, protoplanets become unstable, entering a chaotic evolution that leads to growth via stochastic collisions \citep[e.g.,][]{Lissauer1993}. The detailed effect of such stochastic growth of planets on the overall SFD evolution is unknown, and a much larger number of simulations with larger planetesimal/tracer resolution may be necessary. Despite this, the overall trend of slope evolution seems to be well captured. 

Together, the results presented in Figures \ref{Fig3}, \ref{Fig3Apx}, and \ref{Fig4} demonstrate that terrestrial planet accretion naturally leads to a SFD for leftover planetesimals that is similar in shape to that observed among S-complex bodies in the MAB (Figure \ref{Fig1}). Our results agree with and support the proposition that S-complex asteroids may have formed in the terrestrial planet region\footnote{Assuming, of course, that implantation from the terrestrial planet region into the MAB is not size-dependent, a fact that we consider reasonable given that implantation only occurs after gas disk dispersal.}, i.e., within 1.5 au \citep{Bottke2006,Raymond2017b}. 

Our results are also in agreement with the idea that asteroid (4) Vesta may have been a leftover planetesimal from the tererstrial zone implanted into the inner part of the MAB \citep[$a<$ 2.5 au;][]{Deienno2024}\footnote{See also \citet{March2021,Marchi2022,Zhu2021} and additional discussion on Section \ref{sec:conclusions}.}, where implantation efficiencies are higher \citep{Bottke2006,Izidoro2024}, rather than an object that formed in-situ. On that note, our leftover planetesimal SFDs would suggest that  implantation efficiencies into the MAB have to be smaller than $\mathcal{O}$(10$^{\rm -2}$--10$^{\rm -4}$) to avoid injecting too many Vesta-sized objects or larger into the MAB. 
The capture efficiencies suggested above are in good agreement with values reported by \citet{Bottke2006}, \citet{Raymond2017b}, \citet{Avdellidou2024}, and \citet{Izidoro2024}.

Finally, it is important to note that our results do not necessarily mean that all of the S-complex population in the MAB had to come from the terrestrial planet region. The best we can say is that our results are consistent with the currently observed S-complex SFD.  Our work does not address the amount of material that may or may not have been placed into the MAB \citep{Izidoro2024}. It is possible that the primordial MAB region, although originally depleted in mass \citep{Izidoro2022}, was not empty of planetesimals \citep{Raymond2017b}. Instead, it probably had an upper limit of about 0.2 M$_{\rm Moon}$ in the form of planetesimals \citep{Deienno2024}.

\section{summary and conclusions} \label{sec:conclusions}

The main goal of this work is to investigate whether the population of S-complex MAB could have originated in the terrestrial planet region, i.e., from within about 1.5 au \citep{Raymond2017b,Izidoro2024}. This led us to track the evolution of leftover planetesimals during the accretion of the terrestrial planets from an annulus \citep{Hansen2009}, as predicted and supported by models of solar system formation from rings of planetesimals \citep[see Figure \ref{Fig2};][]{Izidoro2022}. 
We used the code LIPAD \citep{Levison2012} to self-consistently account for the growth and fragmentation of our planetesimal population during terrestrial planet formation \citep[e.g., as in][]{Walsh2016,Walsh2019,Deienno2019}. 
This was done by considering several different parameters (Section \ref{sec:methods}), but most importantly by varying the initial surface density of planetesimals (${\rm \Sigma_{planetesimal} \propto r^{-\gamma}}$ with $\gamma= 1$ and 5.5), the decay timescale for the protoplanetary disk nebular gas (${\rm \tau_{gas}}=$ 0.5, 1, and 2 Myr), and the power law slope of the initial planetesimals' SFD (${\rm N(>D) \propto D^{-q}}$ with $q = 0$ \citep[D$_{0}$ = 100 km in ${\rm N(>D) = D_{0}D^{-q}}$;][]{Klahr2020,Klahr2021}, 3.5 and 5 for 100 km $<$ D $<$ 1000 km \citep[e.g.][]{Youdin2005,Levison2015b}).

We compared the evolved SFD of our leftover planetesimal population at the end of the gas nebula phase with the SFD of observed S-complex asteroids in the MAB (Section \ref{sec:results}; see also Section \ref{sec:intro} and Figure \ref{Fig1} for our definition of S-complex population). We found that both these SFDs have similar shapes in the diameter range $100 < D < 400$~km \citep[our target range -- see Section \ref{sec:intro} and Figures \ref{Fig1}, \ref{Fig3}, and \ref{Fig3Apx};][]{Bottke2005,Bottke2015,Bottke2020,Morbidelli2009}. 

We further investigated possible changes in our leftover planetesimal SFDs by extending two representative cases out to 100 Myr (Figure \ref{Fig4}). Our results show that the slope of our leftover planetesimal population in the size range of interest does not significantly change during this stage.  Instead, it keeps a similar shape to that of the S-complex population. Although we moslty focused on the distribution of objects with D $>$ 100 km, our leftover planetesimal SFDs (Figures \ref{Fig3} and \ref{Fig3Apx}) for D $<$ 100 km are also suggestive of agreement with what is needed to reproduce the crater records on the terrestrial planets, including Earth's Moon, as a consequence of early bombardment from leftover planetesimals \citep[see also \citet{Minton2015} and \citet{Brasser2020crater}]{Bottke2007,Nesvorny2023,Nesvorny2024}.

Our results demonstrate that planetesimal/asteroid implantation from the terrestrial region is consistent with the currently observed MAB S-complex SFD. This statement is valid provided that capture into the MAB is not a size-dependent process. Therefore, S-complex asteroids may indeed be objects that formed in the terrestrial planet region, i.e., within 1.5 au. This may include large asteroids like (4) Vesta.  It was recently proposed that (4) Vesta is potentially a leftover planetesimal implanted into the inner part of the MAB \citep[$a<$ 2.5 au;][see also \citet{March2021,Marchi2022,Zhu2021}]{Deienno2024}, where implantation efficiencies are higher \citep{Izidoro2024}, rather than an object that formed in-situ. 

With that said, we cannot use our work to evaluate the amount of mass that may or may not be implanted into the MAB. The primordial amount of mass that existed in the MAB was the subject of a work conducted by \citet{Deienno2024}, whose used (4) Vesta and current MAB observations as constraints.  They estimated that the MAB primordial mass needs to be smaller than about one fifth of a lunar mass (i.e., $\lesssim$ 2.14$\times$10$^{\rm -3}$ M$_{\oplus}$). Yet, our leftover planetesimal SFDs do suggest that implantation efficiencies have to be smaller than $\mathcal{O}$(10$^{\rm -2}$--10$^{\rm -4}$) to avoid over-implantation of Vesta-sized objects or larger.

Finally, it is important to have in mind the fact that some isotopic differences and similarities, as well as water mass
fractions and oxidation states exist among Earth, Mars, Vesta, and other meteorite groups \citep[e.g.,][]{Yamakawa2010,Schiller2018,Mah2022}. These topics go
beyond the scope of this work and what we can infer from our methods and
data directly. Therefore, although those are important issues to consider, we leave the investigation of what the real implications from meteorite constraints are to future studies. Still, we note that some of those works predictions seem to be in line with both the idea that the solar system formed from rings of planetesimals \citep{Lichtenberg2021,Morbidelli2022,Izidoro2022}, and to the fact that the MAB primordial mass was always very small \citep[though not necessarily null;][]{Deienno2024,Brasser2025}, and that some S-complex asteroids could have been implanted from the terrestrial planet region while others (mostly smaller; D $\lesssim$ 200 km) may have grown in-situ \citep{Deienno2024}.
For example, there is nothing that prevents Ureilites from forming at around 2.6--2.8
au \citep{Desch2018} from a primordial MAB with 0.2 lunar masses \citep{Deienno2024}. Similarly, the isotopic differences between Earth and Mars in the framework of terrestrial planets forming from a
dense ring of planetesimals were addressed and discussed in \citet[see their Fig. 3 and also recent work by \citet{Dale2025}]{Izidoro2022}.

\begin{acknowledgments}

We thank the anonymous reviewer for their valuable and constructive suggestions that helped improving this work. The work of R.D. was partially supported by the NASA Emerging Worlds program, grant 80NSSC21K0387 and the SSERVI CLOE project. A.I. acknowledges EW program. D.N., W.F.B., and S.M. were supported by the SSERVI CLOE project. F.R. acknowledges support from CNPq grant 312429/2023-1.

\end{acknowledgments}

\appendix
\counterwithin{figure}{section}

\section{Summary of all SFD evolutions}\label{sec:apx}

\begin{figure}[!ht]
    \centering
    \includegraphics[width=\linewidth]{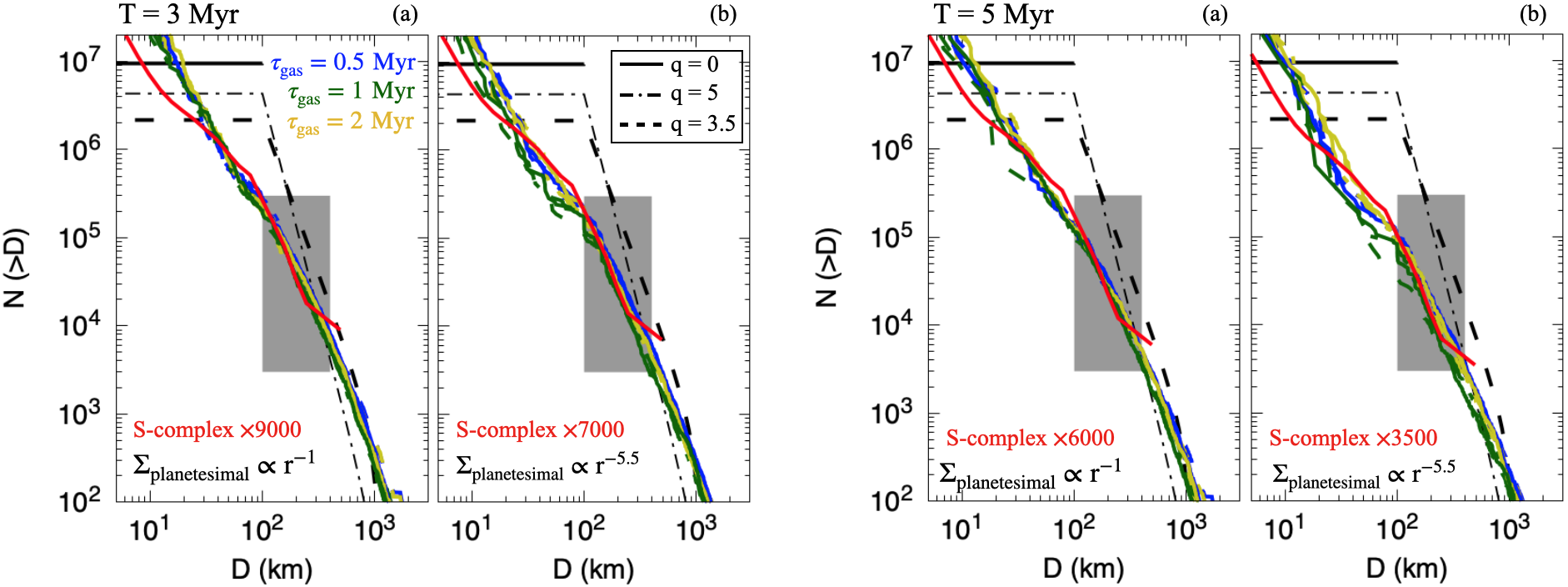}
    \caption{Comparison between leftover planetesimals evolved SFDs in the terrestrial planet region — for different ${\rm \tau_{gas}}$ (blue, green, and yellow) and $q$ (solid, dashed, dot-dashed lines) — with the scaled MAB S-complex SFD (red). Black lines are shown for reference of the initial SFDs. The gray shaded area delimits the diameter range where the leftover planetesimal evolved SFD slopes should be compared to that of the S-complex SFD (Section \ref{sec:intro}, Figure \ref{Fig1}). Left panels show results for T = 3 Myr after CAIs when considering $\gamma=$ 1 (a) and 5.5 (b). Right are the same but for T = 5 Myr after CAIs.}
    \label{Fig3Apx}
\end{figure}

This section is dedicated to present (Figure \ref{Fig3Apx}) a compilation of the results showing all evolved leftover planetesimal SFDs from all 18 simulations performed in this work (i.e., $\gamma=$ 1 and 5.5, ${\rm \tau_{gas}}=$ 0.5, 1, and 2 Myr, with $q$ = 0, 3.5 and 5; Sections \ref{sec:methods} and \ref{sec:results}). Those results, as in Figure \ref{Fig3} were directly compared with the current MAB S-Complex SFD (red in Figure \ref{Fig3Apx}) for objects in the diameter range 100 km $<$ D $<$ 400 km \citep[gray in Figure \ref{Fig3Apx};][]{Bottke2005}. As discussed in Section \ref{sec:methods}, we report our results on the evolved SFDs for both T = 3 and 5 Myr after CAIs due to uncertainties regarding the time when the solar nebula gas fully dispersed in the terrestrial planet region \citep{Weiss2021}. Our SFDs only account for objects smaller than $\approx$ 1000 km, i.e., growing proto-planets (D $\gtrsim$ 1000 km) are not included.

\newpage



\end{document}